\documentstyle{pasj00}

\begin{document}

\SetRunningHead{A.Imada et al.}{Dwarf nova SDSS J013701.06-091234.9}

\title{The 2003/2004 Superoutburst of SDSS J013701.06-091234.9}

\author{Akira \textsc{Imada},$^1$Taichi \textsc{Kato},$^1$Kaori
\textsc{Kubota}$^1$ \\
Makoto \textsc{Uemura},$^2$Ryoko \textsc{Ishioka},$^3$Seiichiro
\textsc{Kiyota},$^4$Kenzo \textsc{Kinugasa}$^5$ \\
Hiroyuki \textsc{Maehara},$^6$Kazuhiro \textsc{Nakajima},$^7$L.A.G. Berto
\textsc{Monard},$^8$Donn \textsc{R.Starkey}$^9$ \\
Arto \textsc{Oksanen},$^{10}$ and Daisaku \textsc{Nogami}$^{11}$}

\affil{$^1$Department of Astronomy, Faculty of Science, Kyoto University,
       Sakyo-ku, Kyoto 606-8502}
\affil{$^2$Hiroshima Astrophysical Science Center, Hiroshima University,
       Hiroshima 739-8526, Japan}
\affil{$^3$Subaru Telescope, National Astronomical Observatory of Japan
       650 North A'ohoku Place, Hilo, HI 96720, U.S.A.}
\affil{$^4$VSOLJ, Center for Balcony Astrophysics, 1-401-810 Azuma,
       Tsukuba, Ibaraki 305-0031}
\affil{$^5$Gunma Astronomical Observatory, 6860-86 Nakayama
       Takayama-mura, Agatsuma-gun, Gunma 377-0702}
\affil{$^6$VSOLJ, Namiki 1-13-4, Kawaguchi, Saitama 332-0034, Japan}
\affil{$^7$VSOLJ, 124 Isatotyo Teradani, Kumano, Mie, Japan}
\affil{$^8$Bronberg Observatory, CBA Pretoria, PO Box 11426, Tiegerpoort 0056, South
       Africa}
\affil{$^9$AAVSO, 2507 County Road 60, Auburn, Auburn, Indiana 46706,
       USA}
\affil{$^{10}$Nyrola Observatory, Jyvaskylan Sirius ry, Kyllikinkatu 1,
       FIN-40100 Jyvaskyla, Finland}
\affil{$^{11}$Hida Observatory, Kyoto University, Kamitakara, Gifu
       506-1314}
\email{a\_imada@kusastro.kyoto-u.ac.jp}

\KeyWords{
          accretion, accretion disks
          --- stars: dwarf novae
          --- stars: individual (SDSS J013701.06-091234.9)
          --- stars: novae, cataclysmic variables
          --- stars: oscillations
}
\maketitle

\begin{abstract}

 We report on time-resolved photometry of the superoutburst of an SU
 UMa-type dwarf nova, SDSS J013701.06-091234.9 in 2003 December-2004
 January. The obtained light curves definitely show superhumps with a
 period of 0.056686 (12) d, which is one of the shortest superhump
 periods among those of SU UMa-type dwarf novae ever
 observed. Considering quiescent photometric studies, we estimated the
 fractional superhump excess to be 0.024. Spectroscopic observations by
 \citet{szk03CVSDSS} provided evidence for TiO bands despite the short
 orbital period, implying that the system has a luminous secondary
 star. We draw a color-color diagram of SU UMa-type dwarf novae in
 quiescence using 2MASS archives, revealing that the location of this
 star in the color-color diagram is deviated from the general trend. The
 distance to the system was roughly estimated to be 300${\pm}$80 pc, using the
 empirical period-absolute magnitude relation and based on the proper
 motion.

\end{abstract}

\section{Introduction}

Dwarf novae are a subclass of cataclysmic variables that consist of a white
 dwarf primary with an accretion disk and a Roche-lobe-filling late type
 secondary star (for a review, see \cite{war95book},
 \cite{hel01book}). 
 
 SU UMa-type stars are a subclass of dwarf novae exhibiting two types of
 outburst: normal outbursts lasting for a few days and superoutbursts
 for weeks, during which $\sim$ 0.2 mag modulations called superhumps are
 always observed. Many models have been proposed in order to explain
 outbursting properties for SU UMa-type dwarf novae. The most promising
 one is the thermal-tidal instability model (for a review, see
 \citet{osa89suuma}) that is also supported by observations.

Most cataclysmic variables below the period gap are believed to evolve
towards short orbital periods due to gravitational-wave
radiation. Minimum of the orbital period of CVs (usually referred to as {\it
period minimum}) has been proposed to be around 65 min
\citep{kol99CVperiodminimum}. It is expected that near the period minimum,
where the secondary begins to degenerate, the inversion of mass-radius
relation causes the orbital period to become longer with evolution. Some
authors claim that WZ Sge-type dwarf novae, which is a subtype of SU
UMa-type dwarf novae, may have experienced the inversion
(\cite{pat05re1255}). It is widely accepted that the most of CVs below
the period gap continue the above mentioned standard evolutional
sequence.

Recently, theory predicts another evolutional sequence for
CVs (\cite{bar00CVsecondary}, \cite{pod03amcvn}). Given a certain range
of mass for the primary and the secondary, the secondary does not become
fully convective in its interior, which leads the secondary to have perpetual
magnetic braking even below the period gap. As a consequence, the
orbital period can become shorter than even the theoretical period
minimum. AM CVn stars \citep{nel05amcvn}, whose orbital periods are less
than 60 min, are the most promising systems that experienced the
aforementioned scenario. EI Psc (= 1RXS J232953.9+062814,
\cite{wei01j2329}, \cite{uem02j2329}, \cite{tho02j2329},
\cite{ski02j2329}, \cite{zho02j2329}) and V485 Cen \citep{ole97v485cen}
may be possible candidates for the progenitor of AM CVn stars.

SDSS J013701.06-091234.9 (hereafter J0137) was first identified as a
cataclysmic variable by \citet{szk03CVSDSS}. Optical spectroscopy for
J0137 showed double-peaked H${\alpha}$ profiles, as well as TiO
bands. Radial velocity studies exhibited a periodicity of 84 min.

An eruption of J0137 had been only recorded by the All Sky Automated
Survey \citep{poj02asas3} as ASAS 013701-0912.6 until the 2003 December
outburst was caught. The first recorded data showed that the object
brightened up to 12.6 in $V$ band on 2001 May 27, then gradually faded
down to 13.4 on June 9, 2001, and finally became below the detection
limit on June 20, 2001. As inferred from the duration of the
brightening, the recorded brightening strongly suggested a superoutburst
of SU UMa-type dwarf novae.

During quiescent photmetric observations of J0137, \citet{pre04j0137}
serendipitously discovered a brightening of the object up to $V=12.5$ mag
on Dec. 21, 2003, when light curves showed superhumps with an amplitude
of 0.2 mag, confirming J0137 as an SU UMa-type dwarf nova.

\section{Observations}

\begin{longtable}{rccccc}
\caption{Journal of observations.}

\hline\hline
Date & HJD(start)$^*$ & HJD(end) & N$^\dagger$ & Exp(s)$^\ddagger$ & Code$^/S$ \\ 
\hline

\hline
\endhead
\hline
\endfoot

23 Dec,2003 & 52996.86904 & 52996.98490 & 151 & 30 & njh \\
 & 52996.88505 & 52997.03497 & 327 & 30 & kyo \\
 & 52997.29186 & 52997.44838 & 444 & 28 & BM \\
24 Dec,2003 & 52997.86596 & 52997.98659 & 159 & 30 & njh \\
 & 52997.93928 & 52998.10752 & 360 & 30 & kis \\
 & 52997.98697 & 52998.11349 & 203 & 30 & gets \\
25 Dec,2003 & 52998.91343 & 52999.05886 & 303 & 30 & kis \\
26 Dec,2003 & 52999.86686 & 53000.00316 & 176 & 30 & njh \\
27 Dec,2003 & 53000.88082 & 53001.08705 & 418 & 30 & kis \\
 & 53000.96131 & 53001.09511 & 225 & 40 & mhh \\
 & 53000.61786 & 53000.69084 & 100 & 45 & DRS \\
28 Dec,2003 & 53002.29974 & 53002.43352 & 375 & 28 & BM \\
29 Dec,2003 & 53002.87155 & 53003.07811 & 359 & 30 & gets \\
30 Dec,2003 & 53003.89347 & 53004.08564 & 322 & 30 & gets \\
 & 53004.25485 & 53004.36336 & 308 & 28 & BM \\
31 Dec,2003 & 53004.95170 & 53005.03557 & 40 & 30 & gets \\
 1 Jan,2004 & 53005.88064 & 53006.05076 & 176 & 30 & gets \\
 & 53006.23946 & 53006.27625 & 48 & 60 & AO \\
 2 Jan,2004 & 53006.95027 & 53007.05426 & 168 & 30 & gets \\
 & 53007.03134 & 53007.05456 & 47 & 30 & kis \\
 3 Jan,2004 & 53007.87306 & 53008.02304 & 241 & 30 & gets \\
 4 Jan,2004 & 53008.96772 & 53009.04763 & 76 & 30 & kis \\
 5 Jan,2004 & 53009.89337 & 53009.91637 & 42 & 30 & gets \\
 & 53009.92619 & 53009.97811 & 105 & 30 & kis \\
 6 Jan,2004 & 53010.92457 & 53010.93495 & 18 & 30 & kis \\
 7 Jan,2004 & 53011.93861 & 53011.97924 & 65 & 30 & kis \\
 8 Jan,2004 & 53012.93191 & 53013.00878 & 96 & 30 & kis \\
10 Jan,2004 & 53014.87110 & 53014.93196 & 48 & 30 & gets \\
12 Jan,2004 & 53016.98983 & 53017.02530 & 37 & 30 & gets \\
13 Jan,2004 & 53017.98725 & 53018.04615 & 92 & 30 & kis \\
14 Jan,2004 & 53018.92030 & 53018.98921 & 105 & 30 & kis \\
15 Jan,2004 & 53019.96453 & - & 1 & 30 & kis \\
\hline
\multicolumn{6}{l}{$^*$ HJD-2400000, $^\dagger$ Number of frames} \\
\multicolumn{6}{l}{$^\ddagger$ Exposure times} \\
\multicolumn{6}{l}{$^\S$ observer's code. (see Table 3.2)} \\
\end{longtable}

CCD photometric observations were performed by the VSNET Collaboration Team
\citep{kat04vsnet}. Table 1 demonstrates log of observations. Used telescopes,
CCDs and sites are listed in Table 2.

\begin{table}
\caption{List of observers.}
\begin{center}
\begin{tabular}{cccc}
\hline\hline
code & site & telescopes & CCD \\
\hline
njh & Mie, Japan & 25cm & CV-04 \\
kyo & Kyoto, Japan & 30cm & ST-7E \\
BM & Pretria, South Africa & 30cm & ST-7E \\
kis & Tsukuba, Japan & 30cm & ST-8E \\
gets & Gunma, Japan & 25cm & AP-7E \\
mhh & Saitama, Japan & 20cm & ST-7E \\
DRS & Indiana, USA & 36cm & ST-10XME \\
AO & Nyrola, Finland & 25cm & ST-8E \\
\hline
\end{tabular}
\end{center}
\end{table}

The Kyoto team used a Java-based point spread function (PSF) photometry
package developed by one of the authors (TK) after dark-subtraction and
flat-fielding. Other observers performed aperture photometry mainly
using AIP4WIN and IRAF package. All CCD systems listed in
Table 1 are close to Kron-Cousins R$_{C}$ band. The magnitudes of each
site were adjusted to the Tsukuba system in which the differential
magnitudes of the variable were measured using TYC2-5277.0337.1 as a
comparison star, whose constancy during the run was used nearby check
stars.

Heliocentric corrections to the observation times were applied before
the following analysis. 

\section{Results}

\begin{figure*}
\begin{center}
\resizebox{160mm}{!}{\includegraphics{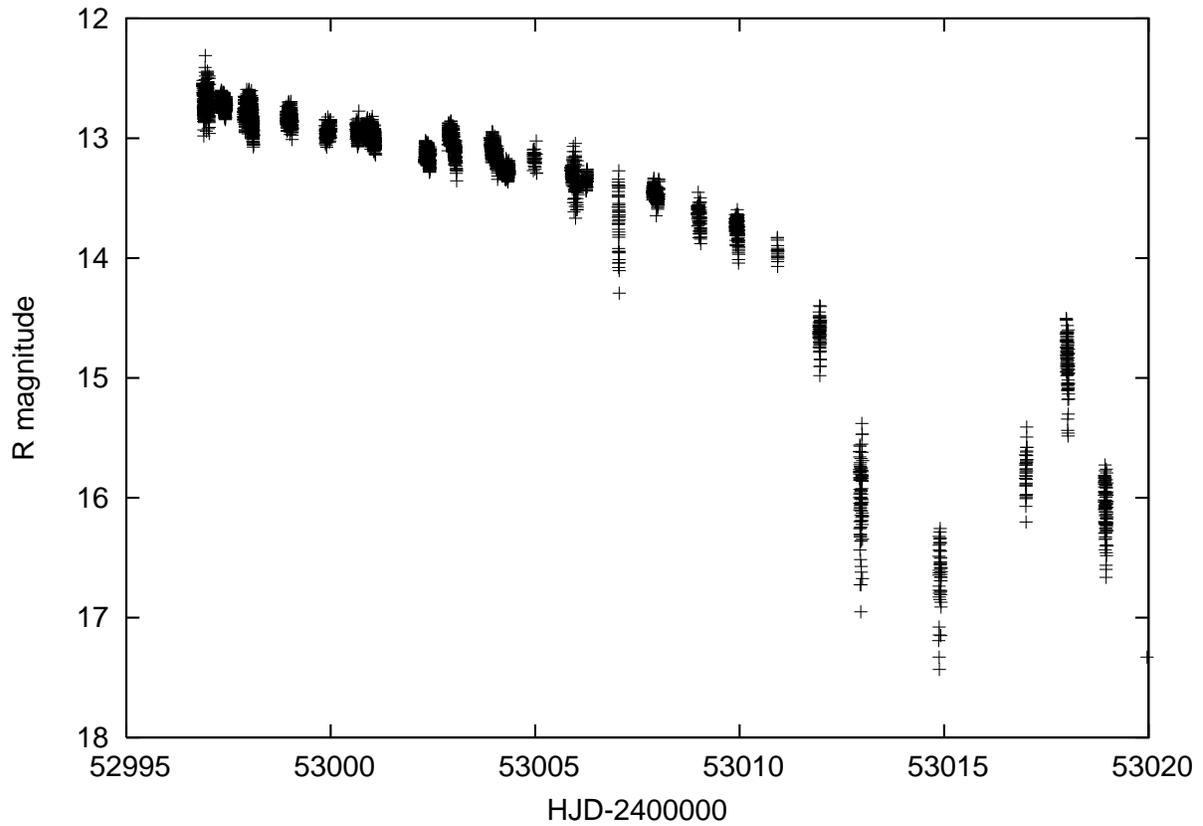}}
\end{center}
\caption{The resulting light curve of SDSS J0137. The abscissa and the
ordinate mean the heliocentric Julian day and the magnitude close to $R$,
respectively. A rebrightening could be seen around HJD 2453018.}
\end{figure*}

\begin{figure*}
\begin{center}
\resizebox{80mm}{!}{\includegraphics{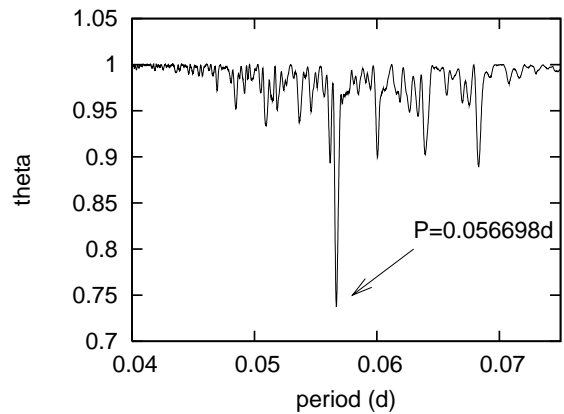}}
\end{center}
\caption{Results of a period analysis by applying the PDM method to the
 data during the plateau stage. The abscissa and the ordinate denote
 periods in the unit of day and theta, respectively. After subtracting
 linear decline trend of the light curve, we determined P = 0.056698
 days as the best-estimated period of superhumps. The second peak close
 to the above derived period seems to be alias. However, we cannot
 exclude the possibility of a real periodicity.}
\end{figure*}

\subsection{Light curve}

Figure 1 represents the obtained light curve of the 2003-2004
superoutburst of J0137. Quiescent photometric observations for the variable
were performed by \citet{pre04j0137} on HJD 2452989 and HJD
2452990, when the object was as faint as $V=18.6$ \citep{pre04j0137}. Then
\citet{pre04j0137} serendipitously detected the outburst of the object
at HJD 2452995.29032, with a magnitude of $V=12.5$. ASAS-3 system also
detected the eruption of the object at HJD 2452993.67590 with a
magnitude of $V=12.8$, while on HJD 2452991.64767 the magnitude was
below the detection limit. Thus the time of the maximum brightness is
restricted to be on HJD 2452994 or HJD 2452995. 

The plateau stage of the superoutburst lasted more than 2 weeks.
The value is slightly longer than that of ordinary SU UMa-type dwarf
novae. Combined the duration of the plateau stage with the fact that the
maplitude of J0137 exceeded 6 mag, the object is within the framework of
large-amplitude SU UMa-type dwarf novae (TOADs, \cite{how95TOAD}). The
mean decline rate during the plateau stage was about 0.08 mag
d$^{-1}$. A rebrightening\footnote{Some authors use the term ``echo
outburst''.} feature is clearly shown on HJD 2453018, which
is often observed in WZ Sge-type dwarf novae and SU UMa-type dwarf novae
with short superhump periods. This implies that J0137 has some relation with
these systems.

We performed a period search of the object during the plateau stage of
the superoutburst after subtracting the linear declining trend. Figure 2
shows the theta diagram using PDM method \citep{ste78pdm}, which
indicates 0.056698 (10) days is the best-estimated superhump period. It
should be noted that the obtained superhump period is comparable to that
of WZ Sge-type dwarf novae, AL Com (0.05722 d, \cite{nog97alcom}), HV
Vir (0.05820 d, \cite{ish03hvvir}), and WZ Sge itself (0.05721 d,
\cite{pat02wzsge}).

\subsection{Superhumps}

\begin{figure*}
\begin{center}
\resizebox{80mm}{!}{\includegraphics{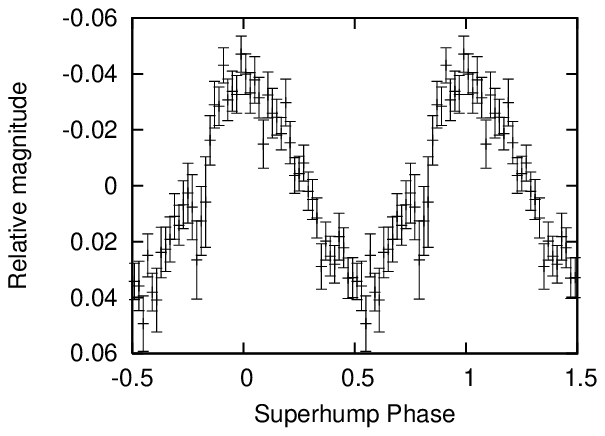}}
\end{center}
\caption{Phase-averaged daily light curves of superhumps during the plateau
stage, after folded by P=0.056698 days. The vertical and the horizontal
 axis denote the relative magnitude and phase, respectively.}
\end{figure*}

\begin{figure*}
\caption{Phase-averaged light curve of superhumps in SDSS
 J013701.06-091234.9, covering between HJD 2452996 and HJD 2453007,
 folded by 0.056698 d. The vertical and the horizontal axes denote the
 relative magnitude and the phase, respectively. A rapid rise and
 slower decline, which is a typical feature of superhumps, are shown
 during the early stage of observations.}
\begin{center}
\resizebox{53mm}{!}{\includegraphics{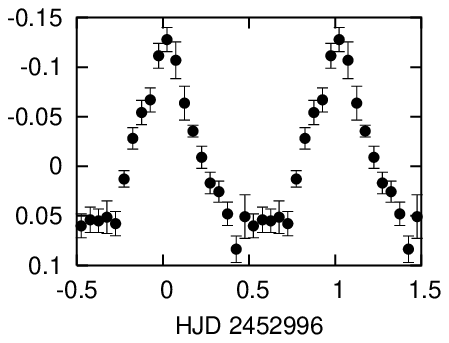}}
\resizebox{53mm}{!}{\includegraphics{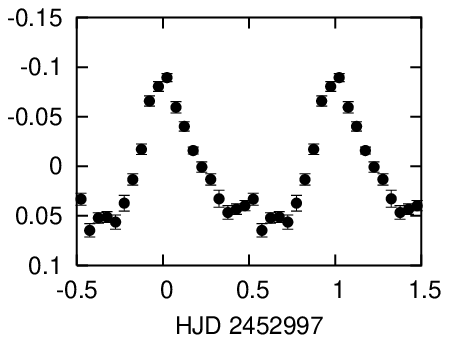}}
\resizebox{53mm}{!}{\includegraphics{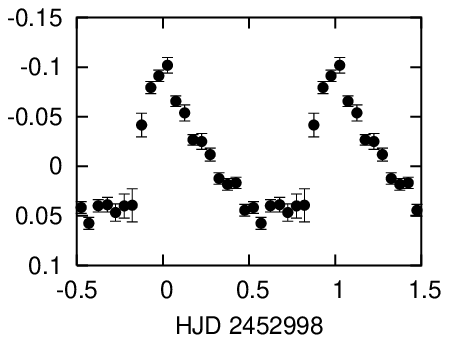}}
\resizebox{53mm}{!}{\includegraphics{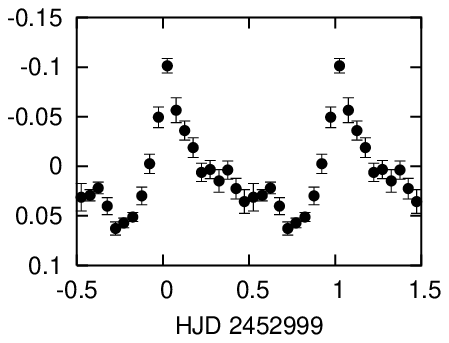}}
\resizebox{53mm}{!}{\includegraphics{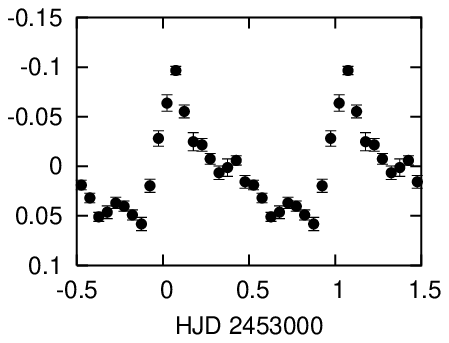}}
\resizebox{53mm}{!}{\includegraphics{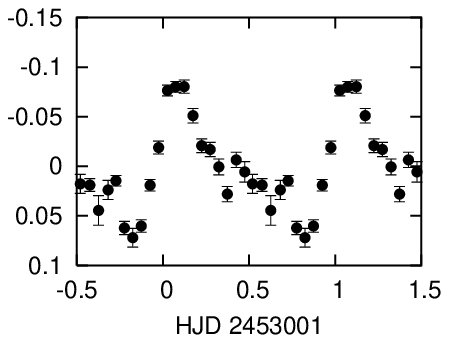}}
\resizebox{53mm}{!}{\includegraphics{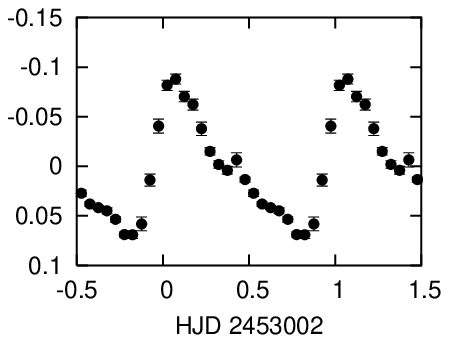}}
\resizebox{53mm}{!}{\includegraphics{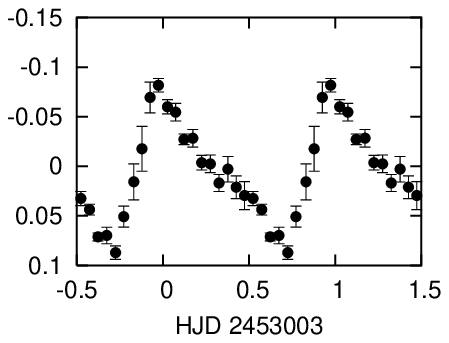}}
\resizebox{53mm}{!}{\includegraphics{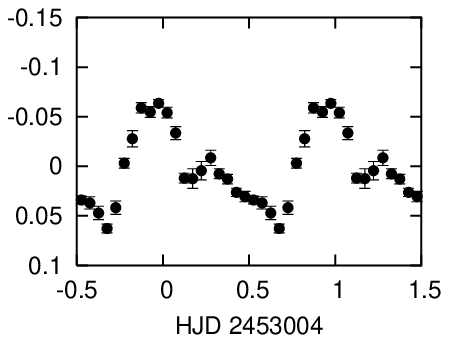}}
\resizebox{53mm}{!}{\includegraphics{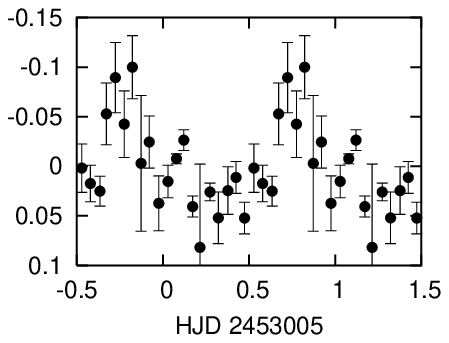}}
\resizebox{53mm}{!}{\includegraphics{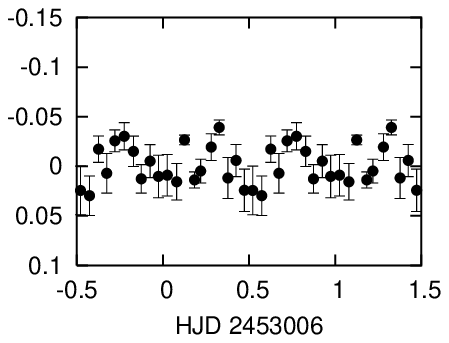}}
\resizebox{53mm}{!}{\includegraphics{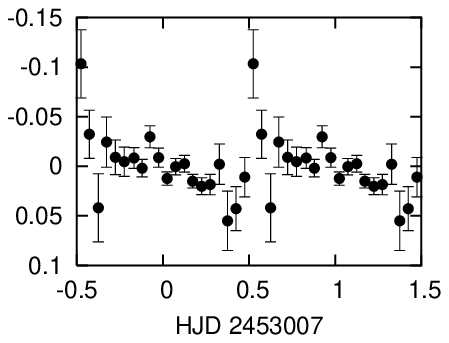}}
\end{center}
\end{figure*}


\begin{figure*}
\begin{center}
\resizebox{80mm}{!}{\includegraphics{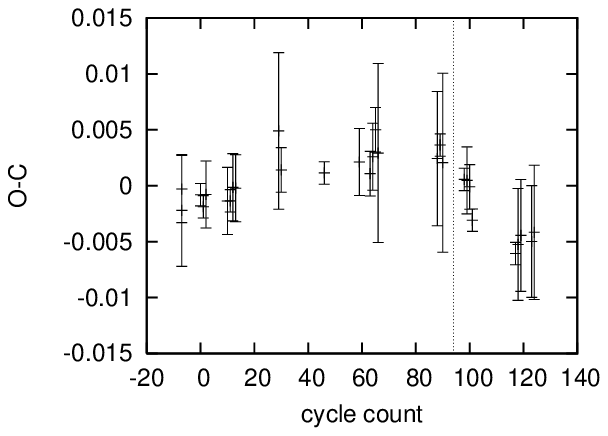}}
\end{center}
\caption{$O-C$ diagram of the superhump maximum timings of J0137 during
 the superoutburst. A calculation is performed based on the equation
 (3). The vertical and the horizontal axes denote $O-C$ and the cycle
 count, respectively. We set $E=0$ to HJD 2452996.9277. Note almost no
 changes of the superhump period till HJD 2453002. The dotted line
 indicates the beginning of late superhumps suggested by
 \citet{pre04j0137}.}
\end{figure*}

The mean superhump profile during the plateau stage is shown in figure
3. A rapid rise and slow decline is typical of that of SU UMa-type
dwarf novae. Figure 4 shows the daily evolution of superhumps during the
plateau stage. Each light curve is folded by the superhump period of
0.056698 d.

We also investigated the existence of early superhumps on HJD
2452996. Early superhumps, having a double-peaked profile and a period
almost the same as their orbital period, are characteristic of WZ
Sge-type dwarf novae (\cite{osa02wzsgehump}, \cite{kat02wzsgeESH},
\cite{pat02wzsge})\footnote{Early superhumps are also called orbital
humps \citep{pat02wzsge} or early humps \citep{osa02wzsgehump}. The
difference among these authors is originated
from their interpretation of physical process near the bright
maximum.}. As can be seen in the top-left panel of figure 4, the
superhumps had already developed on HJD 2452996, which is in agreement
with the observations by \citet{pre04j0137}. However, we cannot exclude
the possibility that early superhumps did emerge near the maximum, and
disappeared before our observations.

\subsection{Superhump period change}

\begin{table}
\caption{Timings of superhump maxima.}
\begin{center}
\begin{tabular}{ccc}
\hline\hline
E & HJD max & error$^\dagger$ \\
\hline
-7 & 52996.9277 & 0.003 \\
0 & 52997.3240 & 0.001 \\
1 & 52997.3796 & 0.001 \\
2 & 52997.4374 & 0.003 \\
10 & 52997.8903 & 0.003 \\
11 & 52997.9470 & 0.001 \\
12 & 52998.0049 & 0.003 \\
13 & 52998.0615 & 0.003 \\
29 & 52998.9736 & 0.007 \\
30 & 52999.0268 & 0.002 \\
46 & 52999.9335 & 0.001 \\
59 & 53000.6714 & 0.003 \\
63 & 53000.8971 & 0.002 \\
64 & 53000.9553 & 0.003 \\
65 & 53001.0144 & 0.002 \\
66 & 53001.0690 & 0.008 \\
88 & 53002.3156 & 0.006 \\
89 & 53002.3735 & 0.001 \\
90 & 53002.4286 & 0.008 \\
98 & 53002.8806 & 0.001 \\
99 & 53002.9372 & 0.003 \\
100 & 53002.9933 & 0.002 \\
101 & 53003.0470 & 0.001 \\
117 & 53003.9510 & 0.001 \\
118 & 53004.0085 & 0.005 \\
119 & 53004.0660 & 0.005 \\
123 & 53004.2922 & 0.005 \\
124 & 53004.3497 & 0.006 \\
\hline
\multicolumn{3}{l}{$^\dagger$ unit in day.}
\end{tabular}
\end{center}
\end{table}

The maximum timings of superhumps measured by eye are listed in Table
3. The typical error is an order of 0.001 d for each maximum. A cycle
count E is set to -7 at the first detected maximum, corresponding to HJD
2452996.9277. A linear regression of the superhump maximum timings
yielded the following equation: 

\begin{equation}
HJD(max)=2452997.3248(9)+0.056686(12){\times}E
\end{equation}

\begin{figure*}
\begin{center}
\resizebox{150mm}{!}{\includegraphics{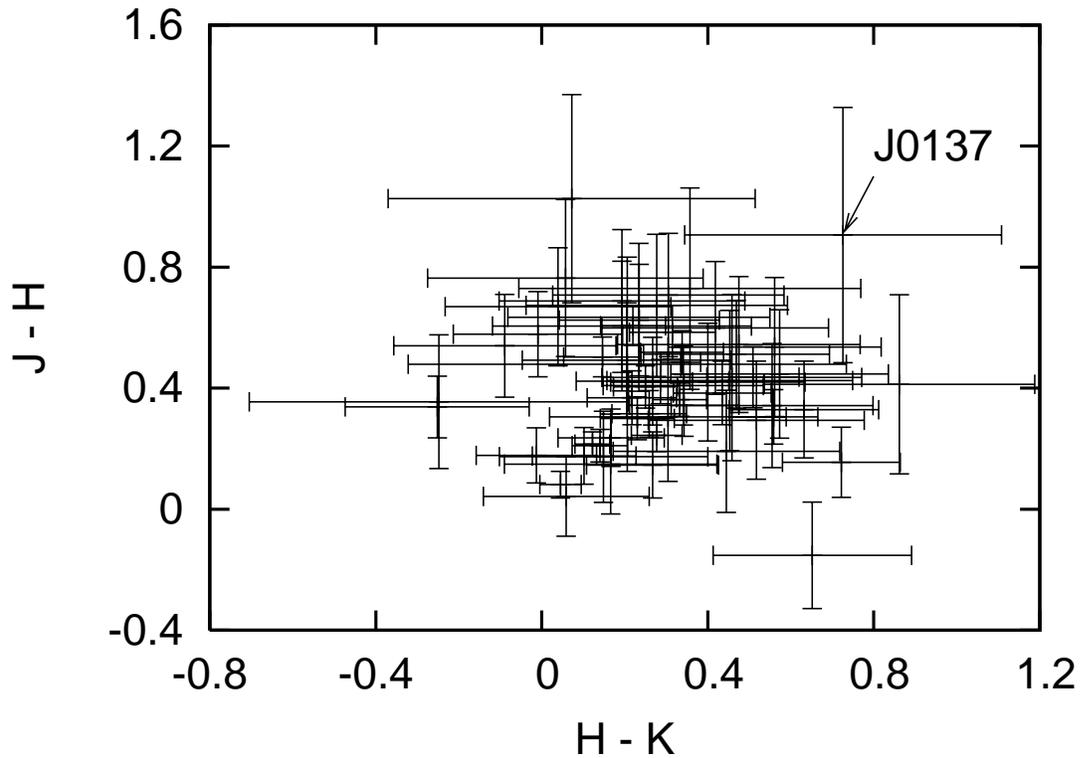}}
\end{center}
\caption{Near-infrared color-color diagram of SU UMa-type stars listed
 on Table 3. Some listed stars incluing ER UMa stars are suspected to be
 observed during outbursts. Such stars for which we cannot evaluate
 typical errors, are precluded. An extinction correction is not
 operated because most of SU UMa-type dwarf novae are not distant and
 the effect is marginal. Note that the location of J0137 is far from
 that of other stars, suggesting the peculiar nature of the object.}
\end{figure*}

The obtained $O-C$ diagram is demonstrated in figure 6, where the
dashed line denotes the beginning of late superhumps discussed in
\citet{pre04j0137}. The data can be apparently fitted by a
quadratic function. However, considering the argument by \citet{pre04j0137}, in
which they have enough data to explore superhump period changes during
the later stage of the outburst, the break in Figure 6 is caused by a
sudden jump of phase as seen in some SU UMa-type dawrf novae. Thus we
conclude that, except the phase change on HJD 2453007, presumably due to
late superhumps, there were almost no changes in superhump periods
during the present superoutburst.

\section{Discussion}

\subsection{outburst properties}

The overall light curves provided firm evidence of superhumps, which
allows us to identify the object as a new SU UMa-type dwarf nova. The
obtained superhump period is 0.056686 (12) d, one of the shortest superhump
periods among SU UMa-type dwarf novae ever known. The plateau stage
lasted more than 2 weeks, and a rebrightening took place at least
once after the termination of the plateau stage. The amplitude of the
object exceeded 6 mag, and the light curve of J0137 has similarity to
that of the 1998 superoutburst of WX Cet \citep{kat01wxcet} and that of
1996 superoutburst of SW UMa \citep{nog98swuma} in terms of the decline
rate, a duration of the plateau stage, and an outburst amplitude.
This certifies J0137 as a large amplitude SU UMa-type dwarf nova (TOADs,
\cite{how95TOAD}).

\subsection{superhump period changes}

It had been considered that superhump periods of SU UMa-type dwarf novae
decrease with time. However, since the 1995 superoutburst of AL
Com, some objects have been confirmed to increase the superhump period
with time. Such systems include AL Com \citep{nog97alcom}, V485
Cen \citep{ole97v485cen}, EG Cnc (\cite{pat98egcnc}, \cite{kat04egcnc}), SW UMa
(\cite{sem97swuma}; \cite{nog98swuma}), V1028 Cyg \citep{bab00v1028cyg},
WX Cet \citep{kat01wxcet}, HV Vir (\cite{kat01hvvir};
\cite{ish03hvvir}), RZ Leo \citep{ish01rzleo}, V592 Cas
\citep{kat02v592cas}, WZ Sge \citep{pat02wzsge}, EI Psc
\citep{uem02j2329}, V1141 Aql \citep{ole03v1141aql}, KS UMa
\citep{ole03ksuma}, VW CrB \citep{nog04vwcrb}, TV Crv
\citep{uem05tvcrv}, GO Com \citep{ima05gocom}, and ASAS 002511+121712
(Imada et al. in prep.)\footnote{It also should be noted that TT Boo
shows both decrease and increase of superhump periods
\citep{ole04ttboo}.}. A decrease of superhump period was interpreted
as being due to shrinkage of the disk radius, or a natural consequence
of mass depletion from the disk \citep{osa85SHexcess}. Most of SU
UMa-type dwarf novae having orbital periods below 0.063 d tend to
increase the superhump period as time elapses \citep{ima05gocom}.

J0137, having the orbital period of 0.0553 d, is expected to be a dwarf
nova with positive derivative of the superhump period. However,
our results clearly indicate that the superhump period hardly changed
during the first half of the plateau stage. \citet{kat01hvvir} suggested
that the period increase may be correlated to a low mass ratio and/or a
low mass transfer rate, which is supported by recent observations
\citep{nog03var73dra}. However, EI Psc and RZ Leo obviously violate the
Kato's suggestion (\cite{uem02j2329};\cite{ish01rzleo}); spectroscopic
observations have revealed a relatively high mass transfer rate and a high
mass ratio of these objects.

Recently, \citet{uem05tvcrv} discovered that an SU UMa-type dwarf nova, TV
Crv exhibits two types of the superhump period change: positive $P_{\rm
dot}$= $\dot{P_{\rm sh}}$/$P_{\rm sh}$ 
during the 2001 superoutburst without a precursor and almost no
changes $P_{\rm dot}$ during the 2004 superoutburst with a
precursor. \citet{uem05tvcrv} suggests that this difference mainly
depends on the disk radius before a superoutburst is triggered. At the
beginning of an outburst, if the accretion disk has
large masses beyond the 3:1 resonance radius at which an eccentric mode
originates, the mode sufficiently propagates outward because of the
plenty matter beyond the 3:1 resonance radius, so that the outer region
of the accretion disk becomes more eccentric, leading to the positive
$P_{\rm dot}$.

J0137, when based on the arguments by \citet{uem05tvcrv}, was likely to
have insufficient mass at the ignition of the outburst. As a consequence,
an eccentric mode could not propagate so far as other SU UMa-type dwarf
novae exhibiting positive $P_{\rm dot}$. The validity of Uemura's arguments
should be explored in the future observations for SU UMa-type dwarf
novae with short orbital periods, and should be tested by hydrodynamical
simulations. 

\subsection{distance}

If the orbital period of the system and the maximum $V$ magnitude of a
normal outburst are known, one can roughly estimate the distance to the
object. In this subsection, we try to estimate the distance to J0137 in
the same manner as that used by \citet{kat03hodel}, \citet{nog04qwser},
and \citet{nog04vwcrb}.

An empirical relation derived by \citet{war87CVabsmag} is that the
absolute $V$ magnitude at the maximum is the function of the orbital
period of the object, that is,

\begin{equation}
M_{V} = 5.64 - 0.259P,
\end{equation}

where $P$ is the orbital period in the unit of hour. This relation,
however, can adapt only to a low inclination system, at most, to an
intermediate inclination system. In the case of J0137, we need caution
to use the equation (2). First, although equation (2) should be used to
the maximum magnitude of a normal outburst of SU UMa stars, a normal
outburst has not been observed for J0137. Second, spectroscopic
observations of J0137 \citep{szk03CVSDSS} shows doubly-peaked profile
of H${\alpha}$, implying an intermediate or high inclination. For the
first caution, we used that the magnitude of a normal outburst 
is 0.5${\pm}$0.5 mag fainter than that of a superoutburst. For the
second caution, the absence of an eclipse can rule out the possibility
of a high inclination system of J0137. Thus, substituting 1.3283 (hr)
into P, we obtain $M_{\rm V}$ ${\sim}$ 5.3. Assuming the maximum
magnitude of a normal outburst to be 12.7${\pm}$0.5, we roughly derived
300${\pm}$80 pc as a likely distance.

We further estimated a distance to J0137 from its proper motion. The
proper motion of J0137 is listed in the USNO B1.0 catalog as
($\mu_{\rm RA}$, $\mu_{\rm Dec}$) = ($-36, -50$) in the unit of
mas. Supposing the transverse velocity to be 100km/s \citep{tho03parallax}, the
distance to J0137 is estimated to be about 320 pc, which does not
contradict that derived above.

\subsection{Secondary star}

\begin{longtable}{cccccccc}
\caption{J, H, and K magnitudes of SU UMa stars in the 2MASS catalog.}

\hline\hline
Object$^\S$ &Offset & Period & Jmag (err) & Hmag (err) & Kmag (err) & Type & Remark$^*$ \\
\hline
\endhead
\hline
\endfoot

EI Psc & 1.546 & 0.044567 & 14.684(43) & 14.316(46) & 14.099(62) & SU & - \\
GW Lib & 1.588 & 0.05332 & 16.191(88) & 15.586(126) & 15.393(186) & WZ? & - \\
BW Scl & 3.029 & 0.054323 & 15.835(83) & 15.493(122) & 14.938(121) & WZ? & - \\
DI UMa & 0.640 & 0.054564 & 15.532(59) & 15.262(109) & 15.126(142) & ER & - \\
V844 Her & 1.451 & 0.05464 & 16.763(150) & 16.284(204) & 16.078(324) & SU & - \\
J0137$^{1}$ & 0.700 & 0.05535 & 16.928(224) & 16.022(198) & 15.296(184) & SU & - \\
J1238$^{2}$ & 0.118 & 0.05592 & 16.651(139) & 16.490(238) & 16.424 & SU? & - \\
HS 2331$^{3}$ & 2.335 & 0.056309 & 15.528(67) & 15.199(93) & 14.567(87) & SU & - \\
PU CMa & 1.097 & 0.05669 & 12.486(24) & 12.461(24) & 12.389(24) & SU & 1, 2 \\
WZ Sge & 0.707 & 0.05669 & 14.862(41) & 14.557(49) & 13.998(57) & WZ & - \\
SW UMa & 0.474 & 0.05682 & 15.621(67) & 15.327(128) & 14.810(132) & SU & - \\
J0532$^{4}$ & 0.924 & 0.057 & 15.175(41) & 15.020(75) & 14.298(67) & SU & - \\
ASAS 0025$^{5}$ & 0.387 & 0.05707 & 16.663(138) & 15.934(195) & 15.577(217) & WZ? & 2 \\
CC Scl & 2.458 & 0.0587 & 16.040(72) & 15.849(130) & 15.404(143) & SU? & - \\
KV Dra & 0.585 & 0.05876 & 16.674(133) & 16.170(255) & 17.118 & SU & - \\
T Leo & 2.385 & 0.05882 & 14.771(43) & 14.335(58) & 13.826(53) & SU & - \\
HS 2219$^{6}$ & 0.090 & 0.0599 & 15.976(92) & 15.440(138) & 14.879(119) & SU & - \\
V1040 Cen & 3.244 & 0.0603 & 16.295(126) & 15.791(149) & 15.707 & SU & - \\
AQ Eri & 1.179 & 0.06094 & 16.425(121) & 15.661(138) & 15.604(194) & SU & - \\
MM Sco & 4.856 & 0.06136 & 16.432(151) & 15.651(141) & 16.488 & SU & - \\
RX Vol & 1.461 & 0.06117 & 16.527(131) & 16.165(202) & 15.185 & SU & 2 \\
V4140 Sgr & 0.963 & 0.06143 & 16.637(126) & 16.700 & 15.903 & SU & - \\
V1159 Ori & 0.929 & 0.062178 & 13.817(27) & 13.781(46) & 13.675(50) & ER & - \\
V2051 Oph & 0.917 & 0.062428 & 14.327(33) & 13.872(43) & 13.530(39) & SU & 1 \\
BC UMa & 0.467 & 0.06261 & 16.785(142) & 16.351(245) & 15.559 & SU & - \\
EK TrA & 0.760 & 0.06288 & 16.490(148) & 16.077(148) & 15.215(178) & SU & - \\
OY Car & 1.855 & 0.06312 & 14.953(37) & 14.435(35) & 14.097(65) & SU & - \\
ER UMa & 1.004 & 0.06366 & 13.606(28) & 13.454(33) & 13.495(37) & ER & - \\
CG CMa & 2.444 & 0.0636 & 15.276(45) & 14.938(57) & 15.190(165) & SU? & 2 \\
V436 Cen & 0.993 & 0.06383 & 14.220(28) & 13.858(27) & 13.526(38) & SU & 2 \\
VY Aqr & 0.863 & 0.06450 & 15.278(52) & 14.855(93) & 14.588(91) & SU & 2 \\
UV Per & 0.488 & 0.06490 & 16.468 & 15.726(147) & 15.431(188) & SU & - \\
AK Cnc & 0.735 & 0.0651 & 13.772(28) & 13.755(39) & 13.716(43) & SU & 1 \\
IX Dra & 0.418 & 0.06646 & 16.470(120) & 16.289(224) & 16.003 & ER & - \\
SX LMi & 0.038 & 0.06720 & 15.707(62) & 15.558(103) & 15.392(153) & SU & - \\
SS UMi & 0.454 & 0.06778 & 15.874(92) & 15.519(129) & 15.767(328) & SU & - \\
CY UMa & 0.733 & 0.06795 & 16.012(67) & 15.500(111) & 15.031(113) & SU & - \\
BZ UMa & 0.511 & 0.06799 & 14.824(43) & 14.440(62) & 14.005(56) & SU & - \\ 
KS UMa & 1.657 & 0.06800 & 16.087(95) & 15.640(117) & 15.067(145) & SU & - \\
RZ Sge & 0.741 & 0.06828 & 15.734(87) & 15.314(108) & 14.914(126) & SU & - \\
TY Psc & 0.661 & 0.06833 & 13.226(21) & 13.145(23) & 13.100(27) & SU & - \\
IR Gem & 1.370 & 0.06840 & 15.218(41) & 14.875(61) & 14.532(71) & SU & - \\
V550 Cyg & 1.403 & 0.0689 & 14.592(50) & 14.166(61) & 13.984(69) & SU & 1, 2 \\
V1504 Cyg & 0.275 & 0.06951 & 16.110(89) & 15.912(154) & 15.402 & SU & - \\
FO And & 0.347 & 0.07161 & 15.493(51) & 15.646(125) & 14.994(114) & SU & - \\
VZ Pyx & 0.731 & 0.07332 & 14.186(33) & 13.871(44) & 13.612(46) & SU & - \\
CC Cnc & 0.302 & 0.07352 & 16.517(120) & 16.082(155) & 15.624(158) & SU & - \\
HT Cas & 1.194 & 0.073647 & 14.703(31) & 14.226(40) & 13.843(55) & SU & 1 \\
IY UMa & 1.255 & 0.07391 & 15.725(92) & 15.100(92) & 14.865(101) & SU & - \\
J1556$^{7}$ & 3.487 & 0.07408 & 16.285(106) & 15.741(119) & 15.266(173) & SU & - \\
VW Hyi & 0.311 & 0.07427 & 12.522(24) & 12.037(26) & 11.702(22) & SU & - \\
Z Cha & 1.616 & 0.074499 & 13.968(35) & 13.564(35) & 13.314(42) & SU & - \\
QW Ser & 0.171 & 0.07457 & 16.274(96) & 15.949(157) & 15.392 & SU & - \\
NSV 10934 & 0.858 & 0.07478 & 14.442(41) & 14.206(54) & 14.039(74) & SU & 2 \\
WX Hyi & 1.250 & 0.07481 & 13.482(26) & 13.238(28) & 12.961(33) & SU & - \\
RZ Leo & 0.777 & 0.07604 & 16.338(116) & 15.664(119) & 15.387(196) & WZ & - \\
SU UMa & 0.464 & 0.07635 & 11.777(22) & 11.731(23) & 11.670(21) & SU & 1 \\
J1730$^{8}$ & 2.681 & 0.07653 & 15.284(47) & 15.189(89) & 15.217(177) & SU & 1 \\
HS Vir & 0.859 & 0.07690 & 15.016(41) & 14.870(68) & 14.603(91) & SU & - \\
V503 Cyg & 2.531 & 0.0777 & 16.370(132) & 15.287 & 15.200 & SU & - \\
V660 Her & 1.007 & 0.07826 & 14.386(35) & 14.395(53) & 14.448(106) & SU & 1 \\
CU Vel & 1.990 & 0.07850 & 14.492(32) & 13.989(34) & 13.843(59) & SU & - \\
V630 Cyg & 0.594 & 0.07890 & 14.679(38) & 14.503(56) & 14.401(69) & SU & - \\
J2100$^{9}$ & 0.086 & 0.079 & 16.100(69) & 16.411(188) & 15.907 & SU & - \\
V1113 Cyg & 1.074 & 0.0792 & 15.777(80) & 15.971(219) & 15.070 & SU & 2 \\
BR Lup & 1.624 & 0.07950 & 15.179(54) & 15.137(77) & 15.078(123) & SU & - \\
DH Aql & 0.853 & 0.08003 & 15.932(86) & 15.263(109) & 15.224(163) & SU & 2 \\
J0549$^{10}$ & 0.405 & 0.08022 & 15.619(50) & 15.210(81) & 14.869(112) & SU? & - \\
PV Per & 0.563 & 0.0805 & 15.181(38) & 15.145(71) & 14.952(114) & SU & 1, 2 \\
TU Crt & 0.933 & 0.08209 & 16.225(104) & 15.517(100) & 15.212(179) & SU & - \\
RX Cha & 0.292 & 0.0839 & 16.765(151) & 16.270(182) & 15.339 & SU & 2 \\
TY PsA & 1.161 & 0.0841 & 14.290(30) & 13.869(43) & 13.583(34) & SU & - \\
V877 Ara & 0.555 & 0.08411 & 16.076(97) & 15.617(135) & 16.181 & SU & 2 \\
J2234$^{11}$ & 0.278 & 0.085 & 16.449(92) & 16.024(140) & 15.571(156) & SU? & - \\
HV Aur & 0.870 & 0.08556 & 13.665(23) & 13.455(31) & 13.315(36) & SU & - \\
DV UMa & 0.181 & 0.08585 & 16.894(172) & 15.868(172) & 15.796(270) & SU & - \\
YZ Cnc & 0.359 & 0.08680 & 13.166(21) & 12.951(19) & 12.829(23) & SU & - \\
IR Com & 0.293 & 0.087039 & 15.032(41) & 14.611(53) & 14.582(86) & SU? & 1 \\
V344 Lyr & 0.595 & 0.08760 & 15.605(51) & 15.432(100) & 15.283(151) & SU & - \\
BF Ara & 0.677 & 0.08797 & 14.788(33) & 14.610(58) & 14.623(87) & SU & 2 \\
V452 Cas & 0.591 & 0.08810 & 15.908(75) & 15.603(138) & 15.299(147) & SU & 2 \\
GX Cas & 1.197 & 0.09297 & 16.226(101) & 15.538(135) & 15.345(161) & SU & 2 \\
MN Dra & 0.410 & 0.10424 & 16.140(91) & 15.782(153) & 15.433 & SU & - \\
TU Men & 0.467 & 0.11720 & 14.747(40) & 14.163(49) & 13.848(55) & SU & - \\
V478 Her & 0.442 & 0.12 & 16.047(77) & 15.555(113) & 15.350(139) & SU? &  - \\
VW Vul & 0.348 & 0.1687 & 13.524(26) & 13.274(29) & 13.168(32) & SU? & 1 \\
ES Dra & 0.103 & 0.179 & 15.458(64) & 14.880(77) & 14.889(127) & SU? & - \\
FS And & 0.589 & - & 15.910(66) & 15.684(119) & 15.340(167) & SU? & 1 \\
BZ Cir & 1.261 & - & 15.612(81) & 15.072(89) & 15.161(179) & SU & - \\
V699 Oph & 0.433 & - & 14.176(32) & 13.570(33) & 13.350(45) & SU & - \\
V823 Cyg & 0.286 & - & 15.038(59) & 14.967(98) & 14.742 & SU & 1 \\
QY Per & 0.841 & - & 15.347(45) & 15.425(106) & 15.261(127) & SU & 1 \\
V2527 Oph & 2.899 & - & 13.699(42) & 13.075(45) & 12.881(43) & SU & 1 \\
V405 Vul & 0.699 & - & 14.531(44) & 14.079(47) & 13.883(49) & SU & 1 \\
EF Peg & 2.151 & - & 12.921 & 15.154(267) & 12.607 & SU & - \\
NSV 907 & 1.141 & - & 16.551(104) & 15.917(140) & 15.683(176) & SU & - \\
J2258$^{12}$ & 2.629 & - & 14.288(34) & 13.988(37) & 13.758(52) & SU & - \\
\hline
\multicolumn{8}{l}{$^\S$: $^{1}$SDSS J013701.06-091234.9, $^{2}$SDSS
 J123813.73-033933.0, $^{3}$HS 2331+3905} \\ 
\multicolumn{8}{l}{$^{4}$1RXS J053234+624755, $^{5}$ASAS 002511+1217.2,
 $^{6}$HS 2219+1824} \\ 
\multicolumn{8}{l}{$^{7}$SDSS J155644.24-000950.2, $^{8}$SDSS
 J173008.38+624754.7, $^{9}$SDSS J210014.12+004446.0} \\ 
\multicolumn{8}{l}{$^{10}$CTCV J0549-4921, $^{11}$SDSS
 J223439.93+004127.2, $^{12}$SDSS J225831.18-094931.7}\\ 
\multicolumn{8}{l}{$^*$1: Photometries may be carried out during outbursts.} \\ 
\multicolumn{8}{l}{2: Orbital periods have not been measured. Instead,
 superhump periods are represented.} \\
\end{longtable}

SU UMa stars with short orbital periods, including WZ Sge stars, show
almost no evidence for the secondary star in the optical
spectrum (e.g., \cite{how01llandeferi}). A quiescent spectrum of J0137,
however, clearly exhibits TiO bands around 7000 A \citep{szk03CVSDSS},
suggesting that a contribution of the M-type secondary is significant
even in the optical range. This implies a peculiar nature of the object
when taking into account the orbital period of J0137 is close to the
theoretical period minimum.

In order to quantitatively investigate secondary stars in SU UMa
stars, we extracted infrared magnitudes of SU UMa stars from the 2MASS
catalog (Table 4) (cf. \cite{hoa02CV2MASS}). These magnitudes could reflect on
the secondary star of each system. All of stars listed in Table 4 are
within 5 arcsec from the coodinates listed in
\citet{DownesCVatlas3}. Using this table, we obtained the color-color
diagram of Figure 7. In this figure, ER UMa stars and SU UMa stars which
could be measured during outbursts are precluded. Note that the location
of J0137 is slightly away from the general trend. This star seems to be
``reddest'', and the colors of $J-H$ and $H-K$ of J0137 is consistent
with those of late M-type to L-type stars, or post-AGB stars
\citep{dah02lateM}. This may mean that the seconday star of J0137 is
later than that of other SU UMa stars. In conjunction with the
spectroscopic observations, we propose that there is an evolved and
luminous secondary star in J0137 like EI Psc and V485
Cen\footnote{However, EI Psc in Figure 7 is located in the general
trend, despite detection of the secondary star in the optical
range. Spectroscopic observations also favor a K-type secondary star in
EI Psc \citep{hu98j2329}. The reason for the difference between EI Psc
and J0137 is left as an open problem.}. The exact spectrum type of the
secondary star should be confirmed in the future observations. 

\subsection{evolutional sequence}

As mentioned above, not only theoretical works \citep{pod03amcvn} but
also observations (\cite{uem02j2329}; \cite{ole97v485cen}) have proposed
an evolutional sequence of binary stars that evolve to AM CVn-type
stars. \citet{kat04lland} suggested that an SU UMa-type dwarf nova LL
And has experienced an intermediate evolution between AM CVn-type stars
and the bulk of CVs, judging from an unexpectedly large superhump excess
of 3.5${\%}$.

J0137 has a superhump excess of 2.4${\%}$. Although the value
is smaller than that of LL And, it is larger than that of SW UMa
(2.1${\%}$, \cite{sem97swuma}, \cite{nog98swuma}) and WX Cet (1.8${\%}$,
\cite{kat01wxcet}) despite their longer orbital periods than that of
J0137. Thus, in conjunction with the peculiar properties of the
secondary star, we suggest that an evolutional sequence of J0137 is
slightly deviated from the most of CVs' evolution as proposed by
\citet{kat04lland} for LL And.

\section{Conclusions}

Photometric observations of the superoutburst of J0137 from December
2003 to January 2004 revealed that (1) the mean superhump period is
0.056686 (12) d, which is one of the shortest superhump periods among SU
UMa-type dwarf novae, (2) the amplitude of J0137 during the
superoutburst exceeded 6 mag and the plateau stage of J0137 lasted about
2 weeks, which confirmed J0137 to be a new member of SU UMa-type dwarf
novae with a large amplitude, (3) the changes
of superhump period were hardly observed during the plateau stage, 
but a signature of late superhumps appeared, (4) a distance to J0137 is roughly
estimated to be 320${\pm}$80 pc, based on the measured proper motion for the
object and an empirical relation given by \citet{war87CVabsmag}, (5)
based on the 2 MASS observations, the secondary star of J0137 has much
later spectral type than that of other SU UMa-type dwarf novae,  and (6)
the fractional superhump excess of J0137 is a large value of 0.024 for
its short orbital period, suggesting that an evolutional sequence of
J0137 may be slightly deviated from that of standard CVs. This implies that
J0137 places an missing link between ordinary SU UMa stars and a
progenitor for AM CVn stars.
  
\vskip 3mm

We are grateful to many VSNET, AAVSO and CBA observers who have reported
vital observations. We also thank B. Warner, P. A. Woudt, and
M. L. Pretorius for the prompt report of the outburst. This work is
supported by Grants-in-Aid for the 21st Century COE ``Center for
Diversity and Universality in Physics'' from the Ministry of Education,
Culture, Sports, Science and Technology (MEXT), and also by
Grants-in-Aid from MEXT (No. 13640239, 16340057, 16740121,
17740105).

\end{document}